%% file: emoji-gender.tex
\documentclass[sigconf]{acmart}
\usepackage{booktabs} 
\usepackage{multirow}
\usepackage{flushend}

\begin{document}
\fancyhead{}
\title{Through a Gender Lens: Learning Usage Patterns of Emojis from Large-Scale Android Users}


\author{Zhenpeng Chen$^1$, Xuan Lu$^1$, Wei Ai$^2$, Huoran Li$^1$, Qiaozhu Mei$^2$, Xuanzhe Liu$^1$}\authornote{Corresponding author: xzl@pku.edu.cn.}
\affiliation{%
  \institution{$^1$Key Lab of High-Confidence Software Technology, MoE (Peking University), Beijing, China\\
  $^2$ School of Information, University of Michigan, Ann Arbor, USA
  }
}
\email{{czp,luxuan}@pku.edu.cn, aiwei@umich.edu, lihuoran@pku.edu.cn, qmei@umich.edu, xzl@pku.edu.cn}

\renewcommand{\shortauthors}{Z. Chen et al.}
\newcommand{\emoji}[1]{\includegraphics[width=1em]{emoji_images/#1.png}}

\newcommand{\para}[1]{\smallskip\noindent{\bf {#1}. }}
\newcommand{\eat}[1]{}

\begin{abstract}
\input{section/0.abstract}
\end{abstract}

%
%


\copyrightyear{2018}
\acmYear{2018} 
\setcopyright{iw3c2w3}
\acmConference[WWW 2018]{The 2018 Web Conference}{April 23--27, 2018}{Lyon, France}
\acmBooktitle{WWW 2018: The 2018 Web Conference, April 23--27, 2018, Lyon, France}
\acmPrice{}
\acmDOI{10.1145/3178876.3186157}
\acmISBN{978-1-4503-5639-8/18/04}



\maketitle

\input{section/1.introduction}
\input{section/2.related}
\input{section/3.dataset}

\input{section/4.preference}

\input{section/5.experiment}

\input{section/6.english}
\input{section/7.discussion}
\input{section/8.conclusion}
\input{section/acknowledgment}

\bibliographystyle{ACM-Reference-Format}
\bibliography{emoji-bibliography}

\end{document}

%% file: section/0.abstract.tex
Based on a large data set of emoji using behavior collected from smartphone users over the world, this paper investigates gender-specific usage of emojis. We present various interesting findings that evidence a considerable difference in emoji usage by female and male users. Such a difference is significant not just in a statistical sense; it is sufficient for a machine learning algorithm to accurately infer the gender of a user purely based on the emojis used in their messages. In real world scenarios where gender inference is a necessity, models based on emojis have unique advantages over existing models that are based on textual or contextual information. Emojis not only provide language-independent indicators, but also alleviate the risk of leaking private user information through the analysis of text and metadata.

%% file: section/1.introduction.tex
\section{Introduction}\label{intro}

On April 11, 2015, Andy Murray, a world-wide known tennis player, announced his wedding on Twitter.\footnote{\url{https://twitter.com/andy_murray/status/586811114744320000}, retrieved on February 10, 2018.} Unlike any other formal announcement, the Tweet consists of no word but 51 emojis.

This is just one of the many evidences that emojis have gained incredible popularity in recent years. Compared to traditional information carriers such as words, pictures, or even emoticons, emojis are considered to be both simple and lively, both expressive and compact, making them widely appreciated by Internet users, particularly by those who use smartphones. 
Emojis have also become an attractive new subject for scientific research. Interwoven into our daily communications, emojis are established as a ubiquitous language that bridges users who speak different languages and who are from different countries, cultures, and demographic groups~\cite{Lu:2016}. Various studies have been done to understand the semantics and sentiments of emojis~\cite{Ai:2017,Novak:2015,tigwell2016oh, MillerTCJTH16,Miller17,PohlDR17,BarbieriRS16}, which concluded that emojis present rich and clear meanings and emotions that can be generalized across language barriers. 

Does ubiquity imply equality? Perhaps not. Previous work has also compared the usage of emojis across apps~\cite{Tauch16}, across platforms~\cite{MillerTCJTH16}, and even across cultures~\cite{Lu:2016}. Considerable differences are demonstrated between these groups in their interpretations and preferences of certain emojis. Our work adds to this literature by examining gender specific usage of emojis. 

Why do we care about genders? Identifying the gender differences in user behaviors is always an important topic in user modeling and human computer interaction. For example, studies have demonstrated that difference exists in how females and males use non-verbal cues in face-to-face offline communications~\cite{Ablon:2013,lafrance:1992,hall1995}. Similar difference also frequently presents in online activities~\cite{Wolf00, hwang2014, Tossell12}. Failing to consider this difference would compromise the quality of information services and interfaces provided to Internet users, such as recommender systems, online advertisements, and social networking tools, and in the long run it could result in inequality in expression and access to information. Indeed, many major information systems provide gender-customized services to their users ~\cite{OtterbacherBC17,KharitonovS12}. Even if gender information is not explicitly available, it is not uncommon to infer it from other information of the users, such as what they say and what they do, in order to improve user profiling and provide personalized services~\cite{KarimiWLJS16,YouBSL14,KratkyC16,emnlp/BurgerHKZ11}, although it may be at the risk of privacy concerns~\cite{IoannidisMWBFT14}. 
In the past decades, gender inference is quite hot and has been widely studied in the research communities such as Web minng, human computer interaction, information retrieval, and natural language processing~\cite{KarimiWLJS16,KratkyC16,YouBSL14,KharitonovS12,JansenS10,emnlp/BurgerHKZ11, icwsm/ZamalLR12, emnlp/SapPEKSKUS14, acl/FlekovaCGUP16, Preotiuc-Pietro16, acl/VolkovaB16,conll/JohannsenHS15,emnlp/CiotSR13}. 

In this paper, we make the first effort to study the gender differences in using emojis. We present an empirical study based on the largest data set of gendered usage of emojis to date, 
which contains 134,419 anonymized Android-smartphone users from 183 countries, with self-reported genders, and their 401 million messages collected in three months, in 58 languages. 
A comprehensive statistical analysis is conducted to analyze various aspects of emoji usage. We find that there exist statistically significant differences between female and male users in emoji usage: (1) women are more likely to use emojis than men; (2) men and women have different preferences for emojis, some of which are consistent with the common beliefs of gender differences; (3) men and women have different preferences in using emojis to express sentiments, some of which are surprisingly different from the common beliefs.

These differences are not just significant in a statistical sense. In fact, they are so strong that a machine learning algorithm can be used to infer gender only from the emoji usage without accessing the textual or any other contextual information. Surprisingly, we find that the overall accuracy can reach 81.1\% for all users, regardless of the language they use. The performance of our approach is comparable to the reported accuracy of the state-of-the-art that infers the genders of Twitter and Facebook users by their textual messages in English~\cite{emnlp/SapPEKSKUS14}
, and the performance is generalizable to non-English users. This result again provides strong evidence of emojis being used as a ubiquitous language. Compared to those built on natural language texts, machine learning models built on emojis are not only generalizable across linguistic barriers, but
also more robust to privacy threats. 



To the best of our knowledge, we make the first effort on analyzing gendered usage of emojis at scale. 
The major contributions of this paper are as follows:
\begin{itemize}

\item We describe the largest data set of gendered usage of emojis to date, covering anonymized users with explicit gender labels, in a large diversity of languages, and from many different countries. 
\item We present a comprehensive empirical analysis on gendered usage of emojis and find that the emoji usage presents statistically significant differences between female and male users.
\item We construct an advanced machine learning model for gender inference purely based on the emojis in a user's messages. The derived model can achieve comparable accuracy to models built on natural English text and the performance is generalizable to non-English users. 
\end{itemize}



The rest of this paper is organized as follows. Section~\ref{related} summarizes related literature. Section~\ref{dataset} describes the data set and how the ethical issues are resolved. Section~\ref{preference} investigates gender difference in emoji usage. Section~\ref{inferring} presents machine learning models for gender inference purely based on emojis used in messages. 
Section~\ref{english} compares our emoji-based models with text-based gender inference algorithms. Section~\ref{discussion} and Section~\ref{limitation} discuss the practical implications and the limitations of the study, followed by concluding remarks in Section~\ref{conclusion}.

%% file: section/2.related.tex
\section{Related Work}\label{related}
We start with a summary of the background and relevant literature. 

\para{Emojis}
The prevalence of emojis has been an attractive phenomenon of social innovation and appreciation. Emojis, graphic symbols carrying specific meanings, 
are widely used to represent real objects and express emotions. 
Much research effort has been spent to study the ubiquitous usage of emojis, including their general popularity ~\cite{Ai:2017} and their different usage across apps~\cite{Tauch16}, across platforms~\cite{MillerTCJTH16}, and even across countries and cultures~\cite{Lu:2016}. Moreover, some researchers focus on the functionality of emojis in online text communication. Besides being used as replacements of content words, emojis are also used in non-verbal ways to decorate text, adjust tones, provide additional emotional or situational information, and engage the audience~\cite{Kelly:2015, Cramer:2016, hu2017spice, PohlDR17, hello2017}. 
Pohl \textit{et al.} investigated gender distribution of Twitter users and found more females tweeting with emojis than males~\cite{PohlDR17}. Such primitive findings suggest potential gender difference in emoji usage and possible biases in analyses that have neglected this difference. 
We systematically study the gender specific usage of emojis and provide insights for future emoji analysis to consider the gender difference. 


\para{Gender Difference}
Gender difference is always an important research topic in sociological and psychological studies from which there are many interesting findings. For example, females are evidenced to show a greater number of facial activities than males~\cite{buck:1980,buck:1982} and observers can identify emotional states more accurately from female faces than from male faces~\cite{buck1974sex}. With the advancement of data science methods, these hypotheses and conjectures about gender differences are measured and tested quantitatively through analyzing online behaviors of users at scale. For example, when the ``facial expressions" (emoticons) become popular in text, researchers investigated the relationship between gender and emoticon usage and found superiority of females in using emoticons~\cite{Wolf00, hwang2014, Tossell12}, which verifies the sociological findings about non-verbal expressivity of females~\cite{Ablon:2013,lafrance:1992,hall1995}. In addition, gender is demonstrated to have effect in online communications. Specifically, women are found to prefer to write about personal topics on social media and to use pronouns, emotional words, interjections, and abbreviations, while men tend to write about philosophical topics, use standard dictionary words, proper names, numbers, technology words, and links~\cite{WangBK13, JurgensTJ17}. However, as trending non-verbal cues in computer-mediated communication, emojis have not been studied systematically from the gender perspective before this paper. 

\para{Gender Inference}
In recent years, identifying genders of users from their online activities has been an active research topic, given its considerable value in personalization and recommender systems. The techniques proposed for this purpose utilize various online information about users, such as their screen names~\cite{KarimiWLJS16}, the images they post on social networks~\cite{YouBSL14}, their interaction behaviors~\cite{KratkyC16}, and the textual content they generate~\cite{emnlp/BurgerHKZ11, icwsm/ZamalLR12, emnlp/SapPEKSKUS14, acl/FlekovaCGUP16, Preotiuc-Pietro16, acl/VolkovaB16,conll/JohannsenHS15}. Most of the studies are conducted with texts using various linguistic cues such as word choices, paraphrase choices, emotions, and part-of-speeches. There is also very limited literature concerning non-English languages. Ciot \textit{et al.} ~\cite{emnlp/CiotSR13} attempted to apply existing English-based gender inference models to other languages and found them not working well. One important reason is the complex orthography of some languages such as Japanese. We use emojis, a ubiquitous language used worldwide and across language barriers, as an indicator of gender and compare the performance of emoji-based gender inference with existing text-based methods. 


%% file: section/3.dataset.tex
\section{The Data Set}\label{dataset}

The data we use in this study are collected through the Kika Keyboard,\footnote{https://play.google.com/store/apps/details?id=com.qisiemoji.inputmethod, retrieved on February 10, 2018.} a leading Android input method app on Google Play. With millions of downloads across the world, Kika supports 82 languages and was among the top 25 most downloaded apps on Google Play in 2015. The data set spans from December 4, 2016 to February 28, 2017, covering 134,419 active users with self-reported gender information and their 401 million messages. 

With emoji inputs as a major feature, Kika supports all emojis released by the Unicode Standard.\footnote{\url{http://unicode.org/emoji/charts/full-emoji-list.html}, retrieved on February 10, 2018.} Our statistics show that 1,356 different kinds of emojis are used in this data set, and 83.9\% users have used emojis at least once. 

The data set has three advanced characteristics that have made this study feasible. First, essential meta information, including the genders and countries of users, is voluntarily reported by users. This supports the analysis of gender difference in emoji usage and provides the ground truth for gender inference. Due to the original design of Kika's information collection procedure, we consider only binary genders in this study, i.e., 53\% females and 47\% males in our data set. Second, data of users with 58 languages from 183 countries and regions are collected, which enables a cross-continent study of gendered patterns and makes it possible to evaluate the generality of our gender-inference approach across multiple languages. Third, because the input method runs at the system level, Kika collects timestamped messages from a wide range of apps, not limiting to the well-studied apps such as Twitter~\cite{Barbieri:2016, emnlp/BurgerHKZ11, icwsm/ZamalLR12}. This enables a more comprehensive analysis of emoji usage instead of being limited to context of social media. 

Note that although the data set contains textual content, it is only used in two ways. One is to infer the language so that we can compare the performance of gender inference for users of different languages (see Section~\ref{inferring}). The other is to reproduce the state-of-the-art text-based gender inference model in order to compare its performance with the proposed emoji-based model (see Section~\ref{english}). 


$\bullet$ \textbf{User Privacy and Ethical Consideration}.
The original data were collected by Kika for the purpose of improving user experience, with explicit user agreements and a strict policy of data collection, transmission, and storage. In this study, we take careful steps to protect user privacy and preserve the ethics of research. First, our work is approved by the Research Ethical Committee of the institutes (a.k.a, institutional review board, or IRB) of the authors. Second, the data set was completely anonymized 
by Kika before provided to the authors. Third, the data are stored and processed on a private, HIPPA-compliant cloud server, with strict access authorized by Kika. The whole process is compliant with the public privacy policy of the Kika company and the best known practice of data mining research.

%% file: section/4.preference.tex
\section{Gender Difference in Emoji Usage}\label{preference}
Previous studies have pointed out that 
females tend to be more non-verbally expressive than males~\cite{Ablon:2013,lafrance:1992,hall1995}. Researchers have also examined the gender difference in the usage of emoticons, precursors of emojis~\cite{Dresner:2010,Tossell12}. 
In this section, we examine how emojis, a new type of non-verbal cues, are used by females and males. 
We start by comparing how frequently people use emojis.  

\subsection{Emoji Popularity}\label{frequency}
As mentioned in Section~\ref{dataset}, emojis are popularly used in our data set. Are emojis equally popular among female and male users? We measure the popularity by calculating the percentage of messages containing emojis (\%emoji-msg). In general, 7.02\% of messages sent by male users contain at least one emoji, while the percentage is 7.96\% for female users, suggesting that females are more likely to use emojis (\emph{p}-value$\ll$0.01, $z$-test~\cite{ztest}).

\begin{figure}
\includegraphics[width=0.7\columnwidth]{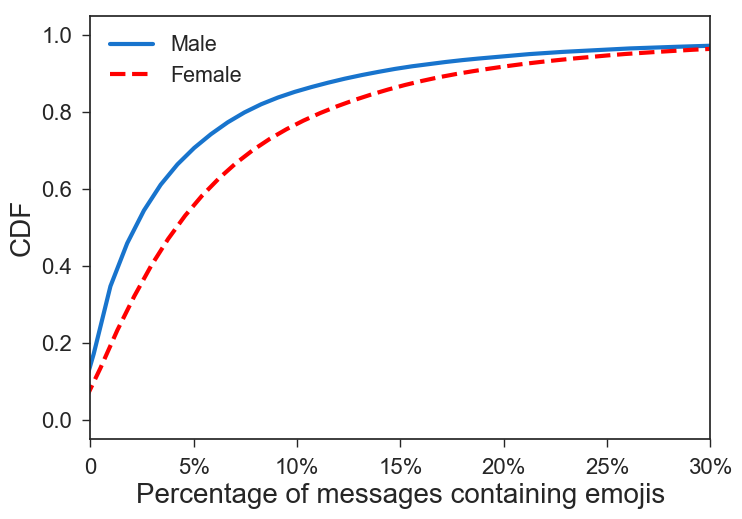}
\caption{Female users include emojis in a larger proportion of their messages.}\label{fig:cdf_emojimsg}
\end{figure}

To further understand the difference, we plot the cumulative distribution function (CDF) of the percentage of users by \%emoji-msg for females and males, respectively. As demonstrated in Figure~\ref{fig:cdf_emojimsg}, the CDF curves are both smooth while the female curve is flatter, which indicates a higher proportion of female users tend to include emojis in more messages. For example, 29.2\% of male users use emojis in more than 5\% of their messages, while the percentage of female users achieves 43.9\%.

To ensure the robustness of our result, we break our data set into three months and compare the \%emoji-msg for females and males in each month. The CDF curves in all three months show that female users have significantly higher tendency to use emojis.

\subsection{Emoji Preference}\label{choice}
The difference in \%emoji-msg, however, does not tell us whether female and male users use different emojis. Do women and men have different preferences for certain emojis? Below we compare the choice of emojis from different genders.

\begin{figure}[!tp]
   \centering
   \includegraphics[width=1\columnwidth]{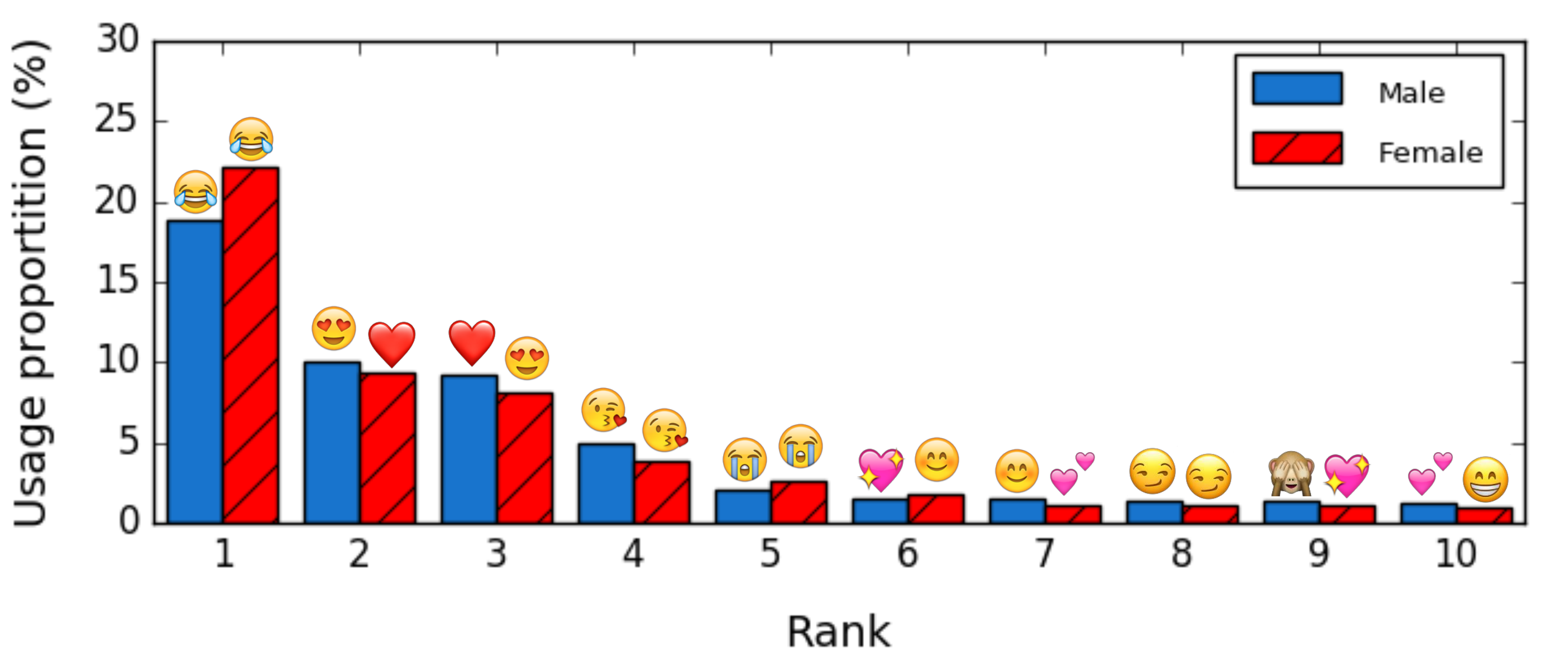}
   \caption{Ten most used emojis by female and male users.}\label{fig:topemoji}
\end{figure}

$\bullet$ \textbf{Frequently Used Emojis}.
We start by comparing the go-to emojis, namely the most frequently used emojis by female and male users. As in Figure~\ref{fig:topemoji}, the emojis used by female or male users both follow a long-tail distribution. 
The 10 most used emojis by women are \emoji{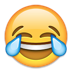}, \emoji{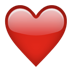}, \emoji{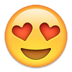}, \emoji{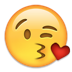}, \emoji{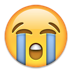}, \emoji{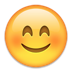}, \emoji{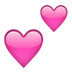}, \emoji{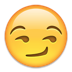}, \emoji{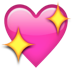}, and \emoji{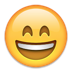}, while the 10 most used emojis by men are \emoji{832}, \emoji{144}, \emoji{843}, \emoji{854}, \emoji{875}, \emoji{592}, \emoji{840}, \emoji{845}, \emoji{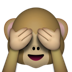}, and \emoji{591}. Interestingly, female and male users have an overlap of 8 emojis in their 10 most-used emojis. 

Beyond the similarities, however, at least two interesting differences can be observed from the two distributions. First, the most popular emoji \emoji{832} (face with tears of joy) accounts for 18.9\% of male users' emoji usage, but 22.1\% for female users. The difference of 3.2\% is non-negligible, as it is even higher than the usage proportion of the 5th most popular emoji for both women and men, \emoji{875} (loudly crying face). The 
difference in favoring \emoji{832} results in a more skewed distribution of emojis used by female users. 
Second, although most of the favored emojis are the same, the rankings of \emoji{144} (red heart), \emoji{843} (smiling face with heart-eyes), \emoji{592} (sparkling heart), \emoji{840} (smiling face with smiling eyes), and \emoji{591} (two hearts) are 
different between the two genders. As expressing sentiment is an important intention of using emojis~\cite{hu2017spice}, the difference in the distributions of top emojis suggests that men and women may convey their sentiments in different ways and we will discuss it in Section~\ref{sentiment}.

$\bullet$ \textbf{Discriminative Emojis}.
From the 6th most popular emoji, we start to see female and male users having different preferences for emojis. We need a more rigorous way to compare their choices on the less popular emojis. More specifically, can we find emojis that are strongly associated with either gender?


To answer this question, we use the \emph{Mutual Information} (MI), which measures the mutual dependence between the usage of a certain emoji and the genders. Emojis with higher MIs are more informative in distinguishing women from men or vice versa. 


Let $Y\in\{1, 0\}$ denote the gender of a user (0 for female and 1 for male).  Let $X\in\{1, 0\}$ denote whether an emoji is used ($x=1$) or not ($x=0$) by this user. The MI for each emoji $e$ can be computed as

\begin{equation*}
\mbox{MI}(X;Y)_{e}=\sum_{x\in X}\sum_{y\in Y}p(x,y)_{e}\log\frac{p(x,y)_{e}}{p(x)_{e}p(y)_{e}},
\end{equation*}
where $p(x)_{e}$, $p(y)_{e}$ are the marginal probabilities of \emph{x} and \emph{y}, and $p(x,y)_{e}$ is the joint probability of \emph{x} and \emph{y}. For example, $p(0,0)_{e}$ is the probability that the emoji $e$ is never used by a male user.

Table~\ref{tab:informative_emoji} lists emojis with the highest MIs, including \emoji{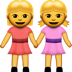} (two women holding hands), \emoji{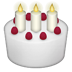} (birthday cake), \emoji{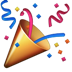} (party popper), etc. In addition, for each discriminative emoji $e$, we calculate \emph{p}(\emph{Female}|\emph{e}), the probability that a user of $e$ is female, and \emph{p}(\emph{Male}|\emph{e}), the probability that a user of $e$ is male. 


As mentioned in Section~\ref{dataset}, 53\% users in our data set are females and the other 47\% are males. We define the emoji $e$ as a \emph{male} emoji if $p(Male|e)>0.47$, otherwise a \emph{female} emoji. Statistics show that there are far more \emph{female} emojis than \emph{male} emojis, and the 10 most informative emojis in Table~\ref{tab:informative_emoji} are all \emph{female} emojis; such results are consistent with findings in linguistics literature~\cite{emnlp/BurgerHKZ11}. In other words, one is more likely to be a female if an emoji in Table~\ref{tab:informative_emoji}, such as \emoji{551}, is ever used by this user. 
We also find some \textit{male} emojis such as \emoji{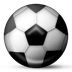} (soccer ball), \emoji{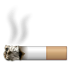} (cigarette), and \emoji{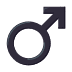} (male sign). By comparison, we find that \textit{female} emojis are fancier and more colorful than \textit{male} emojis, which meets the common interpretations of gender difference. The existence of \emph{female} emojis and \emph{male} emojis evidences the different choice of emojis by female and male users, and suggests the potential of inferring gender through such patterns.
 
\begin{table}[!tp]
  \caption{A selection of discriminative emojis, ranked by mutual information with gender.}
  \label{tab:informative_emoji}
  \begin{tabular}{ccccc}
    \toprule
    Rank & MI & Emoji \emph{e} & \emph{p}(\emph{Male}|\emph{e}) & \emph{p}(\emph{Female}|\emph{e})\\
    \midrule
    1 & 0.0223 & \emoji{551} & 0.126 & \textbf{0.874} \\ 
    2 & 0.0160 & \emoji{324} & 0.236 & \textbf{0.764} \\
    3 & 0.0145 & \emoji{331} & 0.275 & \textbf{0.725} \\
    4 & 0.0139 & \emoji{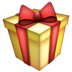} & 0.232 & \textbf{0.768} \\
    5 & 0.0139 & \emoji{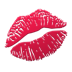} & 0.267 & \textbf{0.733} \\
    6 & 0.0120 & \emoji{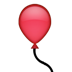} & 0.225 & \textbf{0.775} \\
    7 & 0.0111 & \emoji{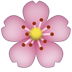} & 0.187 & \textbf{0.813} \\
    8 & 0.0104 & \emoji{591} & 0.310 & \textbf{0.690} \\
    9 & 0.0096 & \emoji{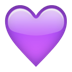} & 0.292 & \textbf{0.708} \\
    10 & 0.0094 & \emoji{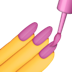} & 0.203 & \textbf{0.797} \\
  \bottomrule
  \end{tabular}
\end{table}

$\bullet$ \textbf{Co-Used Emojis}.
Going one step further, could we compare the context in which the female and male use an emoji? Without messing with different languages, we examine a simple form of context -- the co-used emojis. 
What kind of emojis are frequently co-used by females and males? Is there gender difference in such co-usage patterns? 
To answer such questions, we construct a co-occurrence network for female users and one for male users, respectively. In both networks, the nodes are emojis, and an edge between two emojis is measured by the \textit{Point Mutual Information} (PMI)~\cite{church1990word}, which is formulated as


\begin{equation*}
\mbox{PMI}(e_1,e_2)=\log\frac{p(e_1, e_2)}{p(e_1)p(e_2)},
\end{equation*}
where $p(e_1)$ represents the probability that a message contains $e_1$, $p(e_2)$ represents the probability of $e_2$, and $p(e_1, e_2)$ represents the probability that a message contains both emojis. 

For this network, we connect each emoji to five other emojis that have the largest positive PMI with it, with edges weighted by the corresponding PMI values. By applying the community detection functionality of \emph{Gephi}\footnote{\url{https://gephi.org/}, retrieved on February 10, 2018.} (with resolution as 0.2), we identify 55 communities from the emoji co-occurrence network of female users and 56 communities from the network of male users.\footnote{Due to the random initialization of the algorithm, different runs can produce different communities. Our findings are consistent in different runs.} The nodes within one community have more connections (larger PMI) with each other, while the nodes from different communities have fewer connections (lower PMI).

By comparing communities from the two networks, some interesting findings can be made. For example, we find a sport-related community with emojis like \emoji{95} (soccer ball) and \emoji{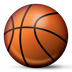} (basketball) from both of the two networks. However, males tend to use such emojis together with \emoji{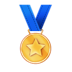} (sports medal) and \emoji{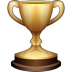} (trophy), while female prefer to use them with \emoji{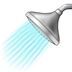} (shower), \emoji{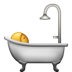} (person taking bath) and \emoji{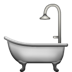} (bathtub). Such a result suggest that females and males may be talking about ``different things'' when they mention sports. Another example is the frequent co-occurrence of clothes-, shoes-, and bag-related emojis with \emoji{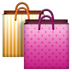} (shopping bags) by female users, which can not be observed from male users. These findings indicate interesting differences in co-used emojis of female and male users.

\subsection{Sentiment Expression}\label{sentiment}
Emojis were originally designed to help express sentiments in a compact and vivid way. Recent research also demonstrates that expressing sentiment is a main intention of using emojis~\cite{hu2017spice}. We infer that the gendered patterns of using emojis (i.e., frequency and preference) can be implicitly affected by the way sentiment is expressed. For example, it is widely believed that women are more emotional and more expressive than men~\cite{buck1974sex}. Can similar observations be made from the sentiments expressed through emojis?

To capture the overall sentiment information, we calculate sentiment scores for each emoji with their official names and annotations from the Unicode Website using LIWC (Linguistic Inquiry and Word Count).\footnote{\url{http://liwc.wpengine.com}, retrieved on February 10, 2018.} With positive (\textit{posemo}) and negative (\textit{negemo}) scores generated by LIWC, each emoji is labelled as \textit{positive} (\textit{posemo}$>$\textit{negemo}), \textit{negative} (\textit{posemo}$<$\textit{negemo}), or \textit{neither}. We calculate the proportions of positive and negative emojis used by female and male users respectively and conduct a \emph{z}-test to measure the difference. Note that when we apply multi-hypothesis tests, we use Bonferroni correction~\cite{dunnett1955} to adjust the $p$-values to obtain more strict and reliable results. Statistics show that female users are more likely to use both positive emojis (female: 50.87\%, male: 50.25\%, \emph{p}-value$\ll$0.01) and negative emojis (female: 10.11\%, male: 9.42\%, \emph{p}-value$\ll$0.01), which is consistent with the existing belief that women are more emotional than men~\cite{buck1974sex}.


In addition to the general proportions, we look at the use of typical emotional emojis, 
i.e., the face-related emojis and heart-related emojis, and explore gender difference in using them. Indeed, we have 69 face-related emojis and 15 heart-related emojis in our data set, and the 84 emojis comprise 75.8\% of total emoji usage for women and 75.5\% for men.

In fact, the two kinds of emojis, the faces and the hearts, perfectly match two typical situations in traditional verbal communication studies. Women are reported to show more facial-related activities than men~\cite{lafrance:1992,buck:1980,buck:1982}, and they are more likely to express love in real life~\cite{fabes:1991,balswick:1971,wilkins:2006}. In textual communications, the face-related emojis emphasize the facial expressions through the eyes, eyebrows, or mouth shapes. Different shapes are used to express different affects and meanings such as happiness (\emoji{840}), depression (\emoji{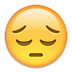}), and anger (\emoji{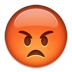}). The heart-related emojis emphasize the color and shape of heart (such as \emoji{144}, \emoji{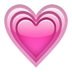}, and \emoji{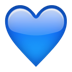}) to convey love and affects in a more direct way. Does emoji usage show similar characteristics with the verbal communication situations? In other words, do female users use more face-emojis and heart-emojis than males? 


We calculate the proportions of face- and heart-related emojis used by female and male users. Face-related emojis are obviously more frequently used than heart-related emojis by both female users and male users. It is understandable because we have far more face-related emojis than heart-related emojis in our data set. By comparing female and male users, we find that female users are significantly more likely to use face-related emojis (female: 58.17\%, male: 56.11\%, \emph{p}-value$\ll$0.01). Such an observation is compliant with previous studies on verbal communication~\cite{lafrance:1992,buck:1980,buck:1982}. However, we are surprised to find that male users are significantly more likely to use heart-related emojis than females (female: 17.62\%, male: 19.41\%, \emph{p}-value$\ll$0.01). This is contrary to psychological literature where males are reported to be less willing to express love in real life~\cite{fabes:1991,balswick:1971,wilkins:2006}. Such a finding implies that although men reserve to express their love in real life, they are more willing to express love through emojis in textual communication. To sum up, women and men have gendered preferences in conveying sentiments through emojis, and some of the preferences are quite different from common interpretations.

%% file: section/5.experiment.tex
\begin{table*}[!tp]
\centering
\caption{Features used to build the emoji-based gender inference model.}
\label{tab:features}
\begin{tabular}{lrp{350pt}}
\hline
Dimension & \# of features & Description\\ \hline
\multirow{5}{*}{Emoji frequency} & \multirow{5}{*}{9} & The overall emoji usage frequency (\%emoji-msg), the average/max/median number of emojis in a message, the proportion of messages using only emojis, the proportion of messages containing only one emoji, the proportion of messages containing multiple nonconsecutive emojis, the proportion of messages containing multiple consecutive emojis, and the proportion of messages containing the same emoji repeatedly.\\
\hline
Emoji preference & 1,356 & The usage proportion of each emoji to all emojis.\\ \hline
\multirow{4}{*}{Sentiment expression} & \multirow{4}{*}{5} & The proportion of positive emojis in total emoji usage, the proportion of negative emojis in total usage, the proportion of messages containing positive emojis, the proportion of messages containing negative emojis, and the proportion of messages containing both positive and negative emojis. \\
\hline
\end{tabular}
\end{table*}

\section{Power of Emojis in Inferring Gender}\label{inferring}
In the previous section, we have compared how women and men use emojis. Not only do they use emoji in different frequencies, but they also have different preferences for selecting which emojis to use. However, how different they are remains a question. In this section, we answer this question by validating the predictive power of emojis. Formally, we attempt to predict the gender of a user purely by their patterns of using emojis.


\subsection{Prediction Set-up}

As we have binary gender labels in our data set (i.e., female and male), we perform gender inference as a binary classification task. We apply several advanced machine models, including the Ridge Classifier (Ridge)~\cite{hoerl1970ridge}, the Random Forest Classifier (RF)~\cite{breiman2001random}, the Gradient Boosting Classifier (GBC)~\cite{friedman2002stochastic}, and the SVM Classifier~\cite{cortes1995support} with a linear kernel. In specific, SVM Classifier with L1 penalization (SVC1) and L2 penalization (SVC2) are used. These algorithms can provide a representative coverage of machine learning methods.

$\bullet$ \textbf{Data Set}.
To obtain robust results, we consider only the users with at least 100 messages that contain emojis. The selected 39,372 users are randomly divided into two subsets, i.e., a training set with 31,872 users and a test set with the other 7,500 users. 

$\bullet$ \textbf{Evaluation Metrics}.
Treating the self-reported gender as the ground-truth, we use the \emph{accuracy} to measure the performance of gender inference models. 
It is measured as the percentage of number-correct over test-size. In addition, we also consider the \emph{precision} of the predicted male and female users, respectively. In specific, we calculate \textit{precision\_M} as the proportion of true male users among those predicted as male, and \textit{precision\_F} as the proportion of true female users among those predicted as female. We adopt these two metrics because in real world applications (such as online advertising~\cite{LiLMWP15}) it is the common practice to provide gender-specific treatments to users of whom the algorithm is confident about the genders and a general treatment to those whose genders are uncertain. It is therefore important for a gender inference algorithm to obtain a high precision. 

$\bullet$ \textbf{Baseline}.
For comparison purposes, we consider a simple baseline according to the gender distribution in the test set. In the test set, we have 4,898 female users and 2,602 male users. Hence, the baseline accuracy is 0.653, the precision of female predictions is 0.653 and the precision of male predictions is 0.347.

\subsection{Feature Extraction}\label{feature}
In Section~\ref{preference}, we have demonstrated the gender difference in three aspects. To train a good model for gender inference, we are inspired to craft 3 sets of features, namely emoji frequency, emoji preference, and sentiment expression. In total we extract 1,370 features for each user, as summarized in Table~\ref{tab:features}. It is true that some features may be collinear with others, but we will leave it to the classification algorithms to handle. 

$\bullet$ \textbf{Emoji Frequency}.
Section~\ref{frequency} has shown that female users are more likely to use emojis in messages, which prompts us to consider the frequency of emojis as features. As we dive into the emojis used in a message, we find that it is not enough to use a single feature \%emoji-msg (i.e., percentage of messages with emoji). For example, people might use multiple emojis in a single message, either by repeating the same emojis like \emoji{834}\emoji{834}\emoji{834} or by rambling different emojis throughout the message. Sometimes emojis constitute the entire message. These patterns may correlate with the intention of why people use emojis~\cite{Tauch16}. To fully capture these patterns about how emojis are used within messages, we construct 9 features. Aside \textit{\%emoji-msg}, the remaining 8 features are as follows.

First, for each message where at least one emoji is used (i.e., an emoji message), we calculate the number of emojis used in it and aggregate the numbers by user. For each user, we calculate \textit{the average/max/median number of emojis in an emoji message}.

Second, we identify the messages where emojis are used in certain patterns, and count the proportion of such messages among emoji messages. These patterns include \textit{emoji only}, \textit{single emoji in text}, \textit{multiple nonconsecutive emojis}, \textit{multiple consecutive emojis}, and \textit{repeating emojis}.

$\bullet$ \textbf{Emoji Preference}.
We have seen in Section~\ref{choice} that women and men have different preferences in choosing which emoji to use. Thus the summary statistics of different emojis used by a user could be discriminative in inferring genders. To this end, we calculate \textit{usage proportion} of each emoji among all emojis typed by the user. The higher the proportion is, the more the user prefers an emoji. In this way, we extract 1,356 features. We have also noticed that women and men have different patterns of emoji co-usage. However, to avoid overfitting, we do not include such patterns in our model. Arguably, including those features could increase the predictive power of emojis.

$\bullet$ \textbf{Sentiment Expression}.
Female and male users can have their own preferences in expressing sentiments through emojis. Based on the results of sentiment classification of emojis (see Section~\ref{sentiment}), we consider the following 5 features.
For each user, we consider \textit{the proportion of positive emojis} and \textit{the proportion of negative emojis} they used, \textit{the proportion of messages containing positive emojis}, \textit{the proportion of messages containing negative emojis}, and \textit{the proportion of messages containing both positive and negative emojis}.

%

\subsection{Model Evaluation}
We use the scikit-learn package in Python~\cite{pedregosa2011scikit} to train the machine learning models, Ridge, RF, GBC, SVC1, and SVC2, using the emoji-based features and the default hyper-parameter settings. In certain situations where we attempt to optimize the hyper-parameters, they are selected using the 5-fold cross validation on the training set by optimizing the accuracy. They include the regularization strength and normalization parameter for Ridge, the number of trees and the maximum depth of the tree for RF and GBC, and the dual and penalty parameter of the error term for SVC1 and SVC2. We report all the test results along with the baseline in Table~\ref{tab:performance}. 

\begin{table}[!tp]
\centering
\caption{Machine learning models trained with default hyper-parameters outperform baseline. Performances with optimal hyper-parameters in parentheses.}
\label{tab:performance}
\begin{tabular}{lrrr}
\hline
\multirow{2}{*}{Model} & \multicolumn{3}{c}{Metrics}\\ \cline{2-4} & Accuracy & Precision\_M & Precision\_F \\ \hline
Ridge & 0.702 (0.718)  & 0.726 (0.702) & 0.699 (0.721) \\
RF & 0.718 (0.758) & 0.702 (0.838) & 0.721 (0.743) \\
GBC & 0.780 (0.811) & 0.769 (0.775) & 0.784 (0.826) \\
SVC1 & 0.731 (0.741) & 0.726 (0.717) & 0.732 (0.747) \\
SVC2 & 0.713 (0.735) & 0.729 (0.714) & 0.711 (0.740) \\
Baseline & 0.653 & 0.347 & 0.653 \\
\hline
\end{tabular}
\end{table}

\begin{table*}[!tp]
\centering
\caption{Model performance in different languages, majority guess baseline in parentheses.}
\label{tab:difflan}
\begin{tabular}{llrrr}
\hline
\multirow{2}{*}{Language} & \multirow{2}{*}{Language Family}  & \multicolumn{3}{c}{Metrics}\\ \cline{3-5} & & Accuracy & Precision\_M & Precision\_F \\ \hline
English & Indo-European & 0.824 (0.684) & 0.744 (0.316) & 0.857 (0.684)\\
Spanish & Indo-European & 0.828 (0.653) & 0.794 (0.347) & 0.843 (0.653) \\
Portuguese & Indo-European & 0.841 (0.665) & 0.825 (0.335) & 0.846 (0.665) \\
Tagalog & Austronesian & 0.793 (0.664) & 0.770 (0.336) & 0.800 (0.664) \\
French & Indo-European & 0.775 (0.645) & 0.727 (0.355) & 0.794 (0.645) \\
Italian & Indo-European & 0.841 (0.661) & 0.793 (0.339) & 0.863 (0.661) \\
Arabic & Afro-Asiatic & 0.764 (0.555) & 0.854 (0.555) & 0.690 (0.445)\\
Indonesian & Austronesian & 0.756 (0.617) & 0.745 (0.383) & 0.760 (0.617) \\
Malay & Austronesian & 0.756 (0.618) & 0.758 (0.382) & 0.756 (0.618) \\
German & Indo-European & 0.783 (0.617) & 0.852 (0.383) & 0.761 (0.617) \\
Thai & Tai-Kadai & 0.808 (0.727) & 0.750 (0.273)& 0.819 (0.727)\\
\hline
\end{tabular}
\end{table*}

All the five algorithms using emoji features outperform the baseline in \textit{accuracy}, \textit{precision\_M}, and \textit{precision\_F} even with default hyper-parameters. 
Such results confirm our assumption that the difference of emoji usage by female and male users is sufficiently large so that it can be utilized to infer the gender.

With optimal hyper-parameters, GBC achieves the best accuracy and precision of female users, and RF obtains the best precision of male users. GBC achieves an accuracy as high as 0.811, which outperforms the baseline by 24\%. 
Additionally, although the gender distribution is quite unbalanced (i.e., 65.3\% female users and 34.7\% male users), the \textit{precision\_M} and \textit{precision\_F} of GBC are quite balanced, indicating a relatively fair prediction model.

\subsection{Generalizing to Specific Languages}
Due to the complexity of natural language processes, existing text based approaches often face the challenge of generalizability across languages. For example, models trained based on English text can hardly generalize to other languages, such as Japanese~\cite{emnlp/CiotSR13}.

The advantage of using emojis is that they can be used in different languages. Now that our model is trained without any textual information, can it be applied to different languages and predict the gender without even knowing a word? Beyond the overall performance, we would like to see how our model perform on each individual language. To answer this question, we identify the languages used by users in the test set with the tool \emph{Language Identification},\footnote{\url{https://pypi.python.org/pypi/langid}, retrieved on February 10, 2018.} and evaluate model performance in different languages.

We select 10 non-English languages with the most users, i.e., Spanish, Portuguese, Tagalog, French, Italian, Arabic, Indonesian, Malay, German, and Thai. The 10 languages cover four language families defined by ISO 639\footnote{\url{https://en.wikipedia.org/wiki/List\_of\_ISO\_639-1\_codes}, retrieved on February 10, 2018.} and can be considered as a reasonable representative of the language systems. Spanish, Portuguese, French, Italian, and German belong to Indo-European; Arabic belongs to Afro-Asiatic; Tagalog, Indonesian, and Malay belong to Austronesian; and Thai belongs to Tai-Kadai. The languages in different language families are genetically unrelated and geographically dispersed. 

We construct 10 test sets with users of a specific non-English language in each set and a test set with only English-speaking users as well. By applying the GBC model with optimal hyper-parameters on the 11 test sets, we find satisfactory \textit{accuracy}, \textit{precision\_M}, and \textit{precision\_F} for all the languages. As demonstrated in Table~\ref{tab:difflan}, our model significantly outperforms the baseline. For example, the \textit{accuracy} in Italian users is 0.841, 27\% higher than the baseline. In addition, the \textit{precision\_M} and \textit{precision\_F} are well balanced in each language. For example, the ratio of men and women in Thai users is 0.273:0.727, while our model achieves a \textit{precision\_M} of 0.750 and a \textit{precision\_F} of 0.819, which further supports the predictive power of emoji patterns for both female and male users in a specific language.

An interesting finding can be observed from Arabic users. Emojis are extremely predictive for male users, for the \textit{precision\_M} is as high as 0.854, 54\% higher than the baseline. However, the \textit{precision\_F} is only 0.690, the lowest in all the 11 languages. A possible explanation might be the cultural effect on self-presentation online~\cite{reed2016thumbs}, such as stricter self-censored behaviors for certain genders.



To sum up, the solid performance of the gender inference model on the 10 non-English languages suggests the advantage of the emoji-based approach over text-based models, the generalizability across languages. Although the predictive power of emoji usage patterns varies in different languages due to complex reasons, most results still support the robustness of the approach.

\subsection{Discussion}
Recall that we selected users who have more than 100 messages containing emojis. We next loosen the restriction in the test set to evaluate the feasibility of our model on relatively ``silent'' users. In specific, we select users with $[$80,100), $[$60,80), $[$40,60), $[$20,40), and $[$1,20) emoji messages to construct test sets and perform the pre-trained GBC model with optimal hyper-parameters on these sets. The number of users in these sets are 4,206, 5,515, 7,511, 12,662, and 43,309, respectively.

\begin{table}[!tp]
\centering
\caption{Prediction performance of users with different numbers of emoji-messages, baseline in parentheses.}
\label{tab:threshold}
\begin{tabular}{lrrr}
\hline
\multirow{2}{*}{\# of emoji-msg} & \multicolumn{3}{c}{Metrics}\\ \cline{2-4}  &Accuracy & Precision\_M & Precision\_F \\ \hline
$[$80,100) & 0.748 (0.618) & 0.709 (0.382) & 0.766 (0.618) \\
$[$60,80) & 0.744 (0.619) & 0.692 (0.381) & 0.770 (0.619) \\
$[$40,60) & 0.712 (0.587) & 0.672 (0.413) & 0.635 (0.587)\\
$[$20,40) & 0.675 (0.555) & 0.635 (0.445) & 0.707 (0.555) \\
$[$1,20) & 0.608 (0.548) & 0.595 (0.452) & 0.664 (0.548)\\
\hline
\end{tabular}
\end{table}

Results are shown in Table~\ref{tab:threshold} with comparison to baseline. Along with the decrease of emoji messages of users, our model shows a slight decrease in \textit{accuracy}. Similar to findings in previous text-based studies~\cite{emnlp/Durme12, emnlp/BurgerHKZ11, emnlp/SapPEKSKUS14}, the fewer messages a user sends, the less accurately one can infer their gender. Nevertheless, the emoji-based model still outperforms the baseline in every user group, suggesting that emoji usage patterns are predictive for genders even for relatively ``silent'' users. 


%% file: section/6.english.tex
\section{Emoji vs. Text}\label{english}
We have demonstrated the predictive power of emoji usage patterns for gender inference by comparing the performance of emoji-based models with a naive baseline, i.e., the majority guess. 
While it is not surprising that emoji-based models outperform the naive baseline, a more interesting question is how they compare with the state-of-the-art text-based models. Most of these models are trained using English messages in social media like Facebook and Twitter. For example, Sap \textit{et al.}~\cite{emnlp/SapPEKSKUS14} derived predictive lexicon for genders from word usage on Facebook, Twitter, and in blogs. This lexicon is reported to achieve an accuracy of 91.9\% on the same type of users. 

In this section, we compare the performance of our emoji-based models with prediction models built based on textual content~\cite{emnlp/SapPEKSKUS14} (referred as \emph{text model}). To construct a comparable data set, we select users that meet the following criteria. First, each user must be an \textbf{English} speaker as identified by the \textit{Language Identification} tool. 
Second, each user must have at least 50 emoji messages in total. Such criteria can make sure that we have enough data for each user. 

We finally select 4,156 users and randomly divide them into the training set (3,306 users) and test set (850 users) with their English messages. In specific, the test set includes 564 female users (66.4\%) and 286 male users (33.6\%). For fair comparison, we implement the following four models and compare their performance:

\begin{itemize}
	\item First, we apply the released model by~\cite{emnlp/SapPEKSKUS14}\footnote{http://www.wwbp.org/lexica.html}  (referred as the \textbf{released text model}) on test users. 
	This model was trained on over ten thousand English users and demonstrated to be well generalizable across social media domains. 
	
	\item Second, to ensure a fair comparison between the \textit{text model} and the \textit{emoji model}, we train both models with the same data. Following the procedure described in~\cite{emnlp/SapPEKSKUS14}, we compute the unigram frequency over the aggregate set of messages from each user as features and apply SVM with a linear kernel and L1 regularization to train the model that only considers text. We name this model the \textbf{retrained text model}.
	
	\item Third, we train the \textbf{emoji model} on the same training set (with the GBC algorithm).
	
	\item Finally, we would like to see if incorporating language-specific semantics helps the emoji model. Specifically, we recompute the sentiment scores of emojis based on their semantic nearest neighbors instead of their official descriptions. Based on our previous work~\cite{Ai:2017}, we apply a state-of-the-art embedding algorithm, LINE~\cite{Tang:2016}, and use its second-order proximity to obtain the nearest neighbor words of each emoji. The new sentiment score is estimated as the mean sentiment scores of the nearest neighbors of the emoji. This model is referred as the \textbf{semantic emoji model}. Notice that the only contribution of textual content is to infer the semantics of emojis, and the gender inference model is still only based on emojis.
\end{itemize}

Except for the released text model, the other three models are all trained using both the default hyper-parameters and optimal hyper-parameters on the training set. The way to optimize the hyper-parameters are the same as described in Section~\ref{inferring}. Test results of the four models are illustrated in Table~\ref{tab:textoremoji}, and we summarize our findings as follows:

\begin{itemize}
\item Released text model vs. retrained text model. Under the default hyper-parameter settings, the released text model can have a high \textit{accuracy} of 0.800, while the result of the retrained text model is only 0.718. One possible reason is that the released text model is built upon a large-scale training corpus which might have a different distribution of the data used to retrain and test the text model.

\item Emoji model vs. retrained text model. When trained on the same data set, the emoji model and the retrained text model can obtain similar accuracy (0.736 vs. 0.718) under the default hyper-parameter settings. The emoji model has a higher \textit{precision\_F}, unless exhaustive hyper-parameter search is applied. In contrast, the retrained text model has a higher \textit{precision\_M}. Such a result suggests that the prediction power of emoji model is comparable to the state-of-the-art text-based model in predicting gender. Hyper-parameter tuning brings in a larger improvement for the text model than for the emoji model, which is reasonable as the text model has a much higher degree of freedom. 

\item Emoji model vs. semantic emoji model. As expected, the semantic emoji model with language-specific semantics performs better. This encouraging result suggests that the performance of the emoji model can be improved with language-specific knowledge (not necessarily the textual content in the message). One may expect that other contexts of the language, such as country and culture, may further improve the emoji model, which we leave for future studies.
\end{itemize}

\begin{table}[!tp]
\small
\centering
\caption{Performance of gender inference models under default hyper-parameter settings, followed by performance with optimal hyper-parameters in parentheses.}
\label{tab:textoremoji}
\begin{tabular}{lrrr}
\hline
\multirow{2}{*}{Model} & \multicolumn{3}{c}{Metrics}\\ \cline{2-4} & Accuracy & Precision\_M & Precision\_F \\ \hline
Released text model & 0.800 & 0.693 & 0.858 \\
Retrained text model & 0.718 (0.855) & 0.871 (0.794) & 0.706 (0.885) \\
Emoji model& 0.736 (0.758) & 0.728 (0.744) & 0.738 (0.761) \\
Semantic emoji model & 0.739 (0.769) & 0.746 (0.747) & 0.738 (0.775)\\
Baseline & 0.664 & 0.336 & 0.664\\
\hline
\end{tabular}
\end{table}

%% file: section/7.discussion.tex
\section{Implications}\label{discussion}
So far, we have found significant gender-specific differences in emoji usage. Utilizing these differences as features, one could infer the gender of a user from their emoji usage, independent to the language they use. We then discuss some practical implications of our findings: how they may benefit real world applications.

\para{Keyboard-layout Adaptation for Emojis}
Based on the gender differences in emoji usage, the most straightforward application of the results is to improve user experience of the current smartphone keyboards. In current OS-native and third-party input method apps (IMA), emojis are always split into multiple categories and displayed in pages. Each page contains some emojis that are displayed in a rather fixed layout. When users want to type in an emoji, they have to swipe left or right to search for it. This approach is hardly user friendly: as the number of emoji grows, users may not be able to fast locate and select the desired emojis. Some efforts have been done to optimize emoji entry speed such as~\cite{DBLPPohlSR16}. Our analysis provides implications for smartphone-side IMA developers, not limited to Kika but also including other keyboard developers or even OS vendors, to optimize their keyboard layout. For example, 
the ranked list of emojis shown on a keyboard's layout should be gender-aware. Additionally, current keyboards can recommend the possible words or emojis that users may type in next. Based on our observations, keyboard developers can improve their algorithms from a gendered perspective. Furthermore, given the fact that emojis are ubiquitously popular even on PC (e.g., the ``touchbar'' feature provided in latest Apple Macbook model supports inputting emojis), the gender-specific usage patterns can be utilized there too. 


\para{Generalized Gender Inference with Low Privacy Risk}
Recently, ``sharing'' user profiles becomes a popular practice between Internet-based applications. For example, users are asked to associate their own social networking accounts with an app that can then acquire their profiles. Indeed, knowing the gender, age, and preferences has lots of technical benefits: user profiling, interface design, personalization, recommender systems, and online advertising. However, as a ``double-edged sword,'' such practice also risks user privacy. In practice, it is always a tradeoff between how much better service a user receives and how much their privacy is violated. 
For example, existing work built machine learning models for gender inference (or user profiling in general) based on text messages of a user and achieved satisfactory accuracy~\cite{emnlp/BurgerHKZ11,icwsm/ZamalLR12,emnlp/SapPEKSKUS14}. Even though user IDs can be anonymized, such NLP techniques are still at the risk of accessing and leaking sensitive, private information of the users that are encoded in free text. For example, among the 7,137 natural language features\footnote{\url{http://www.wwbp.org/lexica.html}, retrieved on February 10, 2018.} extracted for gender reference reported in~\cite{emnlp/SapPEKSKUS14}, 507 of them can be found in the most popular 2,000 first names reported by nameberry\footnote{\url{https://nameberry.com/popular\_names}, retrieved on February 10, 2018.} and 209 of them match the most popular 1,000 surnames reported by mongabay.\footnote{\url{https://names.mongabay.com/data/1000.html}, retrieved on February 10, 2018.} Some of the features also imply that there are other types of sensitive information in the free text messages, such as ``\$''(transactions), ``@yahoo.com'' (email addresses), ``http''(Websites), dates, time, and many numbers (age, phone numbers, personal identifiers, financial information, etc). Apparently, models trained purely based on emojis have much less risk of accessing or leaking private information of users, compared to those based on free text. From our results, it is encouraging that an emoji-based model does not sacrifice the accuracy of gender inference much compared to those based on free text. In scenarios where gender reference is a necessity, a better preservation of privacy is a big win. 
Indeed, an emoji-based model does not compromise the accuracy of gender inference, in many scenarios it may even improve the performance, such as when text content is not available or it is written in different languages.

\section{Limitations}\label{limitation}
One potential limitation of this study is the coverage of our data set. With the current design, Kika only collects self-reported genders with a binary option: female and male. It is not clear whether our results can be generalized to other genders.\footnote{For example, it has been reported that Facebook provides 71 gender options:  \url{https://www.yahoo.com/news/gender-options-facebook-users-064847655.html}, retrieved on February 10, 2018.} There might also exist self-selection biases of Kika users when they report their genders. 

Currently, we focus on the active users in the data set collected by the Kika keyboard. Kika is only one of the third-party input methods supporting emojis in the market, and most popular smartphone manufacturers do support emojis in their built-in input methods. 
Although we don't see any evidence of population bias in Kika users, we are not ruling out the possibility that the results of our study might not generalize to users of other emoji apps. 

Indeed, there are still some confounding factors that can potentially influence our results. For example, previous work evidenced that the differences of emoji renderings across platforms give rise to diverse interpretations of emojis~\cite{MillerTCJTH16, Miller17}, which may influence the user behavior of choosing emojis. Our work targets on only Android users, and we currently cannot demonstrate whether the cross-platform issue matters. We do plan to reproduce our approach on iOS users.

%% file: section/8.conclusion.tex
\section{Conclusion}\label{conclusion}
In this paper, we have presented the first empirical study of emoji usage in smartphone from the gender perspective. Our study is based on a unique and large data set collected by Kika, an input method app. The data set covers 134,419 active users from 183 countries, their self-reported gender information, and their 401 million messages in 58 languages collected in three months. We conduct a multi-dimensional analysis of emoji usage from three aspects, i.e., emoji frequency, emoji preference, and sentiment expression via emojis, and find considerable differences in emoji usage between female and male users. The gender difference is not only statistically significant but also sufficient for accurate gender inference via machine learning algorithms. The gender inference model based on only emoji usage patterns achieves comparable performance to those built upon free text features, and it has a much lower risk of violating user privacy. With the language-independent characteristic, the use of emojis can be a reliable indicator for users in different languages, and the competitive performance of the emoji-based model is generalizable to non-English users.

%% file: section/acknowledgment.tex
\section*{Acknowledgment}
This work was supported by the National Key Research and Development Program under the grant No. 2016YFB1000801, the National Natural Science Foundation of China under grant numbers 61725201 and 61528201. The work of Wei Ai and Qiaozhu Mei was supported by the National Science Foundation under grant numbers 1054199, 1633370, 1131500, and 1620319. The authors would like to appreciate the invaluable supports from Dr. Conglei Yao and Mr. Ning Wang from Kika Tech, and Mr. Sheng Shen and Ms. Haotang Liu from Peking University. Xuanzhe Liu is the corresponding author of this work. 